\documentclass[letterpaper, 10pt, conference]{Classes/ieeeconf}
\IEEEoverridecommandlockouts
\usepackage{Packages/decar-common}
\usepackage{Packages/decar-post}
\usepackage{Packages/myFormat}

\title{\LARGE \bf Frequency-Domain Identification of Passive Discrete-Time Systems via Sequential Convex Programming}

\author{Sepehr Moalemi$^{1}$, Jeff Scruggs$^{2}$ and Peter Seiler$^{1}$%
    \thanks{$^{1}$ S. Moalemi and P. Seiler are with the Department of Electrical and Computer Engineering, and $^{2}$ J. Scruggs is with the Department of Civil and Environmental Engineering, both at the University of Michigan, Ann Arbor, MI 48109, USA.
    {\tt\footnotesize \{moalemi, jscruggs, pseiler\}@umich.edu}.
    }
}

\begin{document}
    \maketitle
    \fontdimen16\textfont2=\fontdimen17\textfont2
    \fontdimen13\textfont2=5pt
    \begin{abstract}
This paper considers identification of passive discrete-time linear time-invariant systems from frequency-response data. A prescribed passivity margin and spectral-radius bound are enforced by a descriptor-form linear matrix inequality that is affine in the system realization matrices and storage certificate. A damped Gauss--Newton sequential convex programming method jointly refines the realization matrices through strongly convex semidefinite subproblems, with Armijo backtracking decreasing the fitting error while preserving both guarantees at every iterate. For unstructured identification, the number of decision variables is reduced without changing the attainable transfer matrices by normalizing the descriptor and certificate matrices and restricting the dynamics matrix to upper Hessenberg form. Numerical results on randomized passive plants show that the proposed method achieves the lowest mean frequency-response fitting error among several passivity-enforcing methods, at both full and reduced model orders.
\end{abstract}

    \section{Introduction}
In many engineering applications, the underlying physical system is known to satisfy an energy-dissipation property. One such property is passivity, under which a system cannot deliver more energy through its ports than it has previously stored or received. Preserving this property supports physically consistent modeling and provides a basis for stability analysis through energy inequalities \cite{Hill_Moylan_QSR}.

Frequency-domain subspace identification provides noniterative estimates of multivariable linear time-invariant (LTI) models directly from sampled frequency responses \cite{mckelvey1996subspace}. In \cite{gumussoy2018improving}, these initial estimates are improved through further iterations using the measured data. The Frequency Domain Subspace Based Identification (FSID) toolbox implements this iterative refinement and supports frequency-dependent weighting~\cite{MCKELVEY2021403}. However, these methods do not generally guarantee that the identified model satisfies input--output properties such as passivity. Even when the true system is passive, limited or noisy data, unmodeled dynamics, and reduced-order modeling can produce a realization that is not. This motivates identification methods that account for both fitting accuracy and passivity enforcement.

One way to enforce passivity is to constrain the identified realization to satisfy the Kalman--Yakubovich--Popov (KYP) inequality~\cite{haddad_Chellaboina}. This condition is convex in the storage certificate for a fixed  model, but is not jointly convex in the realization matrices and certificate. A common workaround is to first identify an unconstrained model and then apply a convex projection or post-processing step that holds some model parameters fixed. {This two-step approach is used in rational macromodeling, where a fitted model is made passive by refitting it with prescribed poles~\cite{coelho2004passive} or by iteratively perturbing its parameters~\cite{grivet2008comparative}. In frequency-domain subspace identification, the state and output matrices are fixed while the input and feedthrough matrices are fit under the KYP inequality~\cite{rodrigues2021novel}.} Since these methods do not optimize all model parameters jointly, the resulting fits can be suboptimal at the prescribed model order.

{Other methods directly enforce passivity during the identification step. Positive real subspace identification imposes a KYP inequality that is convex after a change of variables, but its least-squares objective is weighted by the certificate and is evaluated on time-domain state estimates~\cite{hoagg2004positive_real}. Port-Hamiltonian parametrizations are passive by construction and directly fit the frequency response, but lead to unconstrained nonconvex problems that are solved by quasi-Newton methods~\cite{schwerdtner2021, cherifi2024discrete}. For discrete-time single-input single-output models, positive real vector fitting imposes the KYP inequality in every pole-relocation iteration, with the state and input matrices fixed within each iteration~\cite{rodrigues2024passive_vf}. None of these methods minimizes the frequency-response error over all realization matrices jointly while enforcing passivity by an exact convex constraint.}

{The main contribution of this paper is to develop a sequential convex programming (SCP) method that identifies discrete-time positive real models from frequency-response data by jointly optimizing all realization matrices, with passivity enforced at every iterate by an exact convex constraint. Building on~\cite{revay2024ren}, the method linearizes only the transfer matrix in the least-squares objective and solves a damped Gauss--Newton semidefinite subproblem at each iteration. A shifted KYP certificate enforces a prescribed passivity margin and spectral-radius bound, and a congruence-invariant proximal metric makes each subproblem strongly convex. From a feasible initialization, Armijo backtracking decreases the cost monotonically, and every limit point of the iterates is first-order stationary. The formulation also admits additional affine constraints in prescribed descriptor coordinates. For unstructured identification, normalization of the descriptor and certificate matrices and an upper Hessenberg restriction reduce the number of decision variables without changing the transfer matrices admitted by the shifted certificate condition.}

    \section{Preliminaries and Problem Formulation} \label{sec:preliminaries}
\subsection{Notation}
    A symmetric positive definite matrix is denoted by \(\mbf{A} \succ 0\) and a symmetric negative semidefinite matrix is denoted by \(\mbf{A} \preceq 0\). The spectral radius of a square matrix \(\mbf{A}\) is denoted by \(\rho(\mbf{A})\). The truncated inner product is defined for all \(T \in \mathbb{Z}_{> 0}\) as \(\langle \mbf{u}, \mbf{y}\rangle_T = \sum_{t=0}^{T-1} \mbf{u}^{\trans}_{t} \mbf{y}_{t}\). Define \(\ell_{2e}^{m}\) as the set of all functions \(\mbf{u} : \mathbb{Z}_{\geq 0} \to \mathbb{R}^{m}\) such that \(\langle \mbf{u}, \mbf{u}\rangle_T < \infty\), for all \(T \in \mathbb{Z}_{>0}\). The identity matrix is denoted by \(\eye\){, and \(\mathcal{N}(0, v)\) denotes the normal distribution with mean \(0\) and variance \(v\)}.

\subsection{Positive Real Systems and Passivity}
    Consider a discrete-time LTI system
    \begin{equation}
        \begin{aligned}\label{eq:sys}
            \mbf{x}_{t+1} &= \mbf{A}\mbf{x}_t + \mbf{B}\mbf{u}_t,&
            \mbf{y}_t     &= \mbf{C}\mbf{x}_t + \mbf{D}\mbf{u}_t,
        \end{aligned}
    \end{equation}
    where \(\mbf{x}_t \in \mathbb{R}^{n}\), \(\mbf{u}_t \in \mathbb{R}^{m}\), and \(\mbf{y}_t \in \mathbb{R}^{m}\) are the state, input, and output at time \(t\), respectively. The corresponding transfer matrix is \(\mbf{G}(z) \triangleq \mbf{C}(z\eye-\mbf{A})^{-1}\mbf{B} + \mbf{D}\) for \(z \in \mathbb{C}\) such that \(\det(z\eye-\mbf{A})\neq0\).

    \begin{definition}[Positive Realness \cite{anderson}]\label{def:pr}
        A proper square real-rational transfer matrix \(\mbf{G}(z)\) is \emph{positive real} if it is analytic in \(\lvert z \rvert > 1\) and \(\mbf{G}(z) + \mbf{G}(z)^{\herm} \succeq 0\), for all \(\lvert z \rvert > 1\). For \(\mu \geq 0\), \(\mbf{G}\) is \emph{positive real with margin \(\mu\)} if it is positive real, has no poles on \(\lvert z \rvert = 1\), and \(\mbf{G}(e^{\j\omega}) + \mbf{G}(e^{\j\omega})^{\herm} \succeq \mu\eye\) for all \(\omega \in [0, 2\pi)\).
    \end{definition}

    The following lemma provides a linear matrix inequality (LMI) characterization of positive realness and is a discrete-time version of the KYP lemma.
    \begin{lemma}[{\cite[Lemma~3]{anderson}}] \label{lemma:positive_real}
        A square real-rational transfer matrix \(\mbf{G}(z)\) with minimal realization \((\mbf{A}, \mbf{B}, \mbf{C}, \mbf{D})\) is positive real if and only if there exists a \(\mbf{P} \succ 0\) such that
        \begin{equation}\label{eqn:positive_real}
            \hspace{-5pt}
            \mbf{M}_{\mathrm{KYP}}(\mbs{\theta}) \triangleq
            \begin{bmatrix}
                \mbf{P} - \mbf{A}^{\trans} \mbf{P} \mbf{A}  & \mbf{C}^{\trans} - \mbf{A}^{\trans} \mbf{P} \mbf{B} \\
                \mbf{C} - \mbf{B}^{\trans} \mbf{P} \mbf{A}  &  \mbf{D} + \mbf{D}^{\trans} - \mbf{B}^{\trans} \mbf{P} \mbf{B}
            \end{bmatrix}
            \succeq 0,
        \end{equation}
        where \(\mbs{\theta} \triangleq (\mbf{A}, \mbf{B}, \mbf{C}, \mbf{D}, \mbf{P})\).
    \end{lemma}

    For minimal LTI systems, positive realness of the transfer matrix is equivalent to passivity~\cite[Thm.~13.27]{haddad_Chellaboina}. The following input--output relationship is used to define passivity.
    \begin{definition}[Passivity~{\cite[p.~173]{feedback_systems}}]\label{def:passivity}
        The system represented by the operator \(\mbs{\mathcal{H}}:\ell_{2e}^{m}\to\ell_{2e}^{m}\) is \emph{passive} if \(\exists \beta\in\mathbb{R}_{\leq 0}\) such that \(\langle \mbf{u}, \mbf{y}\rangle_T \geq \beta\) for every input \(\mbf{u}\in\ell_{2e}^{m}\) and every \(T\in\mathbb{Z}_{>0}\), where \(\mbf{y} = \mbs{\mathcal{H}}(\mbf{u})\).
    \end{definition}
\subsection{Problem Formulation}
    Let \(\mbf{Y}_k \in \mathbb{C}^{m \times m}\) be measured frequency-response data of an unknown positive real system at frequencies \(\omega_k\), for \(k = 1, 2, \ldots, N_f\). The objective is to fit this frequency-response data with the transfer matrix \(\mbf{G}(z)\) associated with an LTI realization \((\mbf{A}, \mbf{B}, \mbf{C}, \mbf{D})\) of prescribed order \(n\). The corresponding least-squares cost is defined over realizations satisfying \(\det(e^{\j\omega_k}\eye-\mbf{A})\neq0\) as
    \begin{align}\label{eqn:least_squares}
        J(\mbs{\theta}) \triangleq \sum_{k=1}^{N_f} \norm{\mbf{Y}_k - \mbf{G}(e^{\j\omega_k})}^{2}_{\frob}.
    \end{align}
    The cost is minimized subject to the positive realness constraint in \cref{eqn:positive_real} with \(\mbf{P} \succ 0\). The resulting optimization problem over \(\mbs{\theta}\) has a nonconvex objective and nonconvex constraints.

\subsection{Affine Passivity Constraint via Descriptor Formulation}
    The KYP inequality \cref{eqn:positive_real} is nonconvex due to the multiplicative coupling between the dynamics matrices \((\mbf{A}, \mbf{B})\) and the certificate \(\mbf{P}\). {Sequential methods that linearize this coupling do not guarantee passive iterates and therefore require a correction step after each update~\cite{wang2015accurate}, while convex inner approximations, such as iterative convex overbounding, search only an iterate-dependent subset of the feasible set. Instead,} following the slack-variable formulations in \cite{deOliveiraBernussouGeromel1999,deOliveiraGeromelBernussou2002,revay2024ren}, the descriptor matrix
    \(\mbf{E}\) is introduced to remove this coupling from the passivity condition {without conservatism}. Consider the descriptor realization
    \begin{equation}
        \begin{aligned}\label{eq:desc_sys}
            \mbf{E}\mbf{x}_{t+1} &= \mbf{A}_{\mathrm{d}}\mbf{x}_t + \mbf{B}_{\mathrm{d}}\mbf{u}_t,&%
            \mbf{y}_t            &= \mbf{C}_{\mathrm{d}}\mbf{x}_t + \mbf{D}_{\mathrm{d}}\mbf{u}_t,
        \end{aligned}
    \end{equation}
    with \(\mbf{E}\in \mathbb{R}^{n \times n}\) and the other state matrices of appropriate dimensions. If \(\mbf{E}\) is nonsingular, then \cref{eq:desc_sys} is equivalent to \cref{eq:sys}, with transfer matrix
    \begin{align*}
        \hspace{-4pt}
        \mbf{G}(z) = \mbf{C}_\mathrm{d}(z\mbf{E} - \mbf{A}_{\mathrm{d}})^{-1}\mbf{B}_{\mathrm{d}} + \mbf{D}_\mathrm{d} = \mbf{C}(z\eye - \mbf{A})^{-1}\mbf{B} + \mbf{D},
    \end{align*}
    where \((\mbf{A}, \mbf{B}, \mbf{C}, \mbf{D}) \triangleq (\mbf{E}^{-1}\mbf{A}_{\mathrm{d}}, \mbf{E}^{-1}\mbf{B}_{\mathrm{d}}, \mbf{C}_{\mathrm{d}}, \mbf{D}_{\mathrm{d}})\).

    The following lemma, adapted from \cite[Thm.~3]{revay2024ren}, provides a passivity condition for the descriptor realization \cref{eq:desc_sys} that is affine in \(\mbs{\theta}_{\mathrm{d}} \triangleq (\mbf{E}, \mbf{A}_{\mathrm{d}}, \mbf{B}_{\mathrm{d}}, \mbf{C}_{\mathrm{d}}, \mbf{D}_{\mathrm{d}}, \mbf{P}_{\mathrm{d}})\).
    \begin{lemma}\label{lem:desc_passivity}
        A discrete-time LTI system with a minimal realization \((\mbf{A},\mbf{B},\mbf{C},\mbf{D})\) is passive if and only if it admits a transfer-equivalent descriptor realization \cref{eq:desc_sys} with \(\mbf{E}\) nonsingular and a matrix \(\mbf{P}_{\mathrm{d}}\succ0\) such that
        \begin{align}\label{eq:desc_lmi_plain}
            \mbf{M}(\mbs{\theta}_{\mathrm{d}})
            \triangleq
            \begin{bmatrix}
                \mbf{E}+\mbf{E}^\trans-\mbf{P}_{\mathrm{d}} & \mbf{C}_{\mathrm{d}}^\trans                      & \mbf{A}_{\mathrm{d}}^\trans \\
                \mbf{C}_{\mathrm{d}}                        & \mbf{D}_{\mathrm{d}}+\mbf{D}_{\mathrm{d}}^\trans & \mbf{B}_{\mathrm{d}}^\trans \\
                \mbf{A}_{\mathrm{d}}                        & \mbf{B}_{\mathrm{d}}                             & \mbf{P}_{\mathrm{d}}
            \end{bmatrix}
            \succeq0.
        \end{align}
    \end{lemma}
    \begin{proof}
        For sufficiency, the \((1,1)\) block of \cref{eq:desc_lmi_plain} gives \(\mbf{E} + \mbf{E}^\trans \succeq \mbf{P}_{\mathrm{d}} \succ 0\). Consequently, \(\mbf{E}\) is nonsingular. Pre- and post-multiplying \cref{eq:desc_lmi_plain} by \(\mbs{\zeta}_{t}^{\trans}\triangleq \begin{bmatrix}\mbf{x}_{t}^{\trans} & \mbf{u}_t^{\trans} & -\mleft( \mbf{P}_{\mathrm{d}}^{-1}\mbf{E}\mbf{x}_{t+1} \mright)^{\trans}\end{bmatrix}\) and \(\mbs{\zeta}_t\) gives
        \begin{align}\label{eq:traj_plain}
            \hspace{-6pt}
            \mbf{x}_t^\trans \mleft(\mbf{E} + \mbf{E}^\trans \hspace{-1pt}- \mbf{P}_{\mathrm{d}}\mright) \mbf{x}_t - \mbf{x}_{t+1}^\trans \mbf{E}^\trans \mbf{P}_{\mathrm{d}}^{-1} \mbf{E}\, \mbf{x}_{t+1} + 2\mbf{u}_t^\trans \mbf{y}_t \geq 0.
        \end{align}
        Next, by completing the square
        \begin{align*}
            \mbf{E}^\trans \mbf{P}_{\mathrm{d}}^{-1} \mbf{E} - \mleft(\mbf{E} + \mbf{E}^\trans - \mbf{P}_{\mathrm{d}}\mright) = \mleft(\mbf{E} - \mbf{P}_{\mathrm{d}}\mright)^\trans \mbf{P}_{\mathrm{d}}^{-1} \mleft(\mbf{E} - \mbf{P}_{\mathrm{d}}\mright) \succeq 0,
        \end{align*}
        it follows that
        \begin{align}\label{eqn:bound_E_completion}
            \mbf{x}_t^\trans \mleft(\mbf{E} + \mbf{E}^\trans - \mbf{P}_{\mathrm{d}}\mright) \mbf{x}_t \leq \mbf{x}_t^\trans \mbf{E}^\trans \mbf{P}_{\mathrm{d}}^{-1} \mbf{E}\, \mbf{x}_t.
        \end{align}
        Substituting \cref{eqn:bound_E_completion} into \cref{eq:traj_plain} yields
        \begin{align}\label{eq:diss_plain}
            V(\mbf{x}_{t+1}) - V(\mbf{x}_t) \leq \mbf{u}_t^\trans \mbf{y}_t,
        \end{align}
        where \(V(\mbf{x}) \triangleq \tfrac{1}{2}\mbf{x}^\trans \mbf{E}^\trans \mbf{P}_{\mathrm{d}}^{-1} \mbf{E}\, \mbf{x}\) is positive definite. Summing \cref{eq:diss_plain} from \(t = 0\) to \(T - 1\) yields \(\langle \mbf{u}, \mbf{y}\rangle_T \geq V(\mbf{x}_T) - V(\mbf{x}_0) \geq -V(\mbf{x}_0)\), which proves passivity of \cref{eq:desc_sys}. For necessity, by \Cref{lemma:positive_real}, there exists \(\mbf{P}\succ0\) satisfying \cref{eqn:positive_real}. Choose \(\mbf{E} = \mbf{P}\), \(\mbf{A}_{\mathrm{d}} = \mbf{P}\mbf{A}\), \(\mbf{B}_{\mathrm{d}} = \mbf{P}\mbf{B}\), \(\mbf{C}_{\mathrm{d}} = \mbf{C}\), \(\mbf{D}_{\mathrm{d}} = \mbf{D}\), and \(\mbf{P}_{\mathrm{d}} = \mbf{P}\). This construction preserves the transfer matrix. Moreover, \(\mbf{P}_{\mathrm{d}} = \mbf{P} \succ 0\) and \cref{eqn:positive_real} are equivalent to \cref{eq:desc_lmi_plain}. This follows because the Schur complement of the \((3,3)\) block of the matrix in \cref{eq:desc_lmi_plain} is exactly the KYP matrix in \cref{eqn:positive_real}.
    \end{proof}

    \section{Main Results} \label{sec:main_results}
\subsection{Stability and Passivity Margin}
    Although \Cref{lem:desc_passivity} provides an affine way to impose passivity, the nonstrict LMI in \cref{eq:desc_lmi_plain} permits unit-circle poles, which can render the cost \cref{eqn:least_squares} undefined at sampled frequencies. Moreover, \cref{eq:desc_lmi_plain} enforces only nonstrict passivity and provides no quantitative margin from the boundary of the passive set. Therefore, {two parameters are introduced}, \(\eta\in(0,1)\) to exclude unit-circle poles and \(\mu \geq 0\) to prescribe the desired positive real margin. The resulting condition is the shifted LMI
    \begin{align}\label{eq:desc_lmi}
        \mbf{M}_{\eta, \mu}(\mbs{\theta}_{\mathrm{d}})
        \triangleq
        \begin{bmatrix}
            \mbf{H}               & \mbf{C}_{\mathrm{d}}^\trans                                & \mbf{A}_{\mathrm{d}}^\trans \\
            \mbf{C}_{\mathrm{d}}  & \mbf{D}_{\mathrm{d}}+\mbf{D}_{\mathrm{d}}^\trans - \mu\eye & \mbf{B}_{\mathrm{d}}^\trans \\
            \mbf{A}_{\mathrm{d}}  & \mbf{B}_{\mathrm{d}}                                       & \mbf{Q}
        \end{bmatrix}
        \succeq 0,
    \end{align}
    where \(\mbf{H}\triangleq\mbf{E}+\mbf{E}^{\trans}-(1+\eta)\mbf{P}_{\mathrm d}\) and \(\mbf{Q}\triangleq(1 - \eta)\mbf{P}_{\mathrm{d}}\). Notably, \cref{eq:desc_lmi} is affine in \(\mbs{\theta}_{\mathrm{d}}=(\mbf{E}, \mbf{A}_{\mathrm{d}}, \mbf{B}_{\mathrm{d}}, \mbf{C}_{\mathrm{d}}, \mbf{D}_{\mathrm{d}}, \mbf{P}_{\mathrm{d}})\). The following proposition establishes the stability and passivity guarantees of the shifted LMI\@.
    \begin{proposition}\label{prop:desc_lmi}
        Let \(\eta\in(0,1)\), \(\mu\geq0\), and define
        \begin{align}\label{eq:radius}
            r \triangleq \sqrt{\frac{1-\eta}{1+\eta}} \in (0,1).
        \end{align}
        Suppose \(\mbs{\theta}_{\mathrm{d}}\) satisfies \cref{eq:desc_lmi} with \(\mbf{P}_{\mathrm{d}}\succ0\). Then \(\mbf{E}\) is nonsingular, and the equivalent state-space realization
        \begin{align}\label{eqn:desc_to_std_transfer}
            (\mbf{A},\mbf{B},\mbf{C},\mbf{D})=(\mbf{E}^{-1}\mbf{A}_{\mathrm{d}}, \mbf{E}^{-1}\mbf{B}_{\mathrm{d}}, \mbf{C}_{\mathrm{d}},\mbf{D}_{\mathrm{d}})
        \end{align}
        admits the certificate \(\mbf{P}=\mbf{E}^{\trans}\mleft((1-\eta)\mbf{P}_{\mathrm{d}}\mright)^{-1}\mbf{E}\succ0\) which satisfies
        \begin{align}\label{eq:desc_kyp}
            \mbf{M}_{\mathrm{KYP}}(\mbs{\theta}) \succeq \diag\mleft((1-r^2)\mbf{P},\mu\eye\mright).
        \end{align}
        Consequently, \(\rho(\mbf{A})\leq r<1\), and the corresponding
        transfer matrix is positive real with margin \(\mu\).
    \end{proposition}
    \begin{proof}
    From \cref{eq:desc_lmi}, it follows that \(\mbf{H} \succeq 0\). Thus, \(\mbf{E}\) is nonsingular and  \(\mbf{P}=\mbf{E}^{\trans}\mbf{Q}^{-1}\mbf{E}\succ0\). Taking the Schur complement of \(\mbf{Q}\) in \cref{eq:desc_lmi} and substituting \cref{eqn:desc_to_std_transfer} gives
    \begin{align}\label{eqn:desc_schur_proof}
        \begin{bmatrix}
            \mbf{H} -\mbf{A}^{\trans} \mbf{P} \mbf{A} & \mbf{C}^{\trans} -\mbf{A}^{\trans}\mbf{P}\mbf{B}\\
            \mbf{C} - \mbf{B}^{\trans}\mbf{P}\mbf{A}  & \mbf{D}+\mbf{D}^{\trans}-\mu\eye  -\mbf{B}^{\trans}\mbf{P}\mbf{B}
        \end{bmatrix}
        \succeq 0.
    \end{align}
    Rearranging \cref{eqn:desc_schur_proof} and substituting the KYP matrix \(\mbf{M}_{\mathrm{KYP}}(\mbs{\theta})\) from \cref{eqn:positive_real} gives
    \begin{align}\label{eqn:desc_kyp_b4sub}
        \mbf{M}_{\mathrm{KYP}}(\mbs{\theta}) \succeq \diag\mleft(\mbf{P}-\mbf{H},\mu\eye\mright).
    \end{align}
    By completing the square, it can be shown that
    \begin{align}\label{eqn:desc_kyp_comp_square}
        r^2\mbf{P}-\mbf{H} =\frac{1}{1+\eta}\mleft(\mbf{H}-\mbf{E}^{\trans}\mright)^{\trans}\mbf{P}_{\mathrm d}^{-1}\mleft(\mbf{H}-\mbf{E}^{\trans}\mright) \succeq 0.
    \end{align}
    Therefore, \(\mbf{P}-\mbf{H}\succeq(1-r^2)\mbf{P}\). Substituting this bound into \cref{eqn:desc_kyp_b4sub} proves \cref{eq:desc_kyp}. The \((1,1)\) block of \cref{eq:desc_kyp} gives \(\mbf{A}^{\trans}\mbf{P}\mbf{A}\preceq r^2\mbf{P}\). Therefore, \(\rho(\mbf{A})\leq\norm{\mbf{P}^{1/2}\mbf{A}\mbf{P}^{-1/2}}_2\leq r<1\). Moreover, \(\mbf{F}(z)\triangleq(z\eye-\mbf{A})^{-1}\mbf{B}\) is well-defined for all \(\lvert z\rvert\geq1\). Pre- and post-multiplying \cref{eq:desc_kyp} by \(\begin{bmatrix}\mbf{F}(z)^{\herm}&\eye\end{bmatrix}\) and its conjugate transpose, and using \(\mbf{A}\mbf{F}(z)=z\mbf{F}(z)-\mbf{B}\), gives \(\mbf{G}(z)+\mbf{G}(z)^{\herm}\succeq\mleft(\lvert z\rvert^2-r^2\mright)\mbf{F}(z)^{\herm}\mbf{P}\mbf{F}(z)+\mu\eye\succeq\mu\eye\) for all \(\lvert z\rvert\geq1\). Therefore, \(\mbf{G}\) is positive real with margin \(\mu\).
\end{proof}

\begin{remark}[{Coupling of \(\eta\) and \(\mu\)}]
    {The parameters \(\eta\) and \(\mu\) do not act independently. Note that \cref{eq:desc_kyp} is the KYP inequality \cref{eqn:positive_real} for the realization \((r^{-1}\mbf{A}, r^{-1}\mbf{B}, \mbf{C}, \mbf{D}-\tfrac{\mu}{2}\eye)\) with certificate \(r^2\mbf{P}\). Hence, by \Cref{lemma:positive_real}, a minimal realization admits a certificate \(\mbf{P}\succ0\) satisfying \cref{eq:desc_kyp} if and only if \(\mbf{G}\) is analytic and satisfies \(\mbf{G}(z)+\mbf{G}(z)^{\herm}\succeq\mu\eye\) in \(\lvert z\rvert>r\). The margin \(\mu\) is therefore enforced throughout the region \(\lvert z\rvert>r\), which is more restrictive than imposing \(\rho(\mbf{A})\leq r\) and a margin \(\mu\) on the unit circle separately. This restriction vanishes as \(\eta\to0\).}
\end{remark}

\subsection{Equivalent Passive Descriptor Realizations}
    The descriptor realization in \cref{eq:desc_sys} admits equivalent passive representations under nonsingular changes of state coordinates. More specifically, for any nonsingular \(\mbf{T}\), define
    \begin{align}\label{eqn:congurance}
        \mbs{\theta}_\mbf{T} \triangleq \mleft(\mbf{T}^\trans\mbf{E}\mbf{T},\mbf{T}^\trans\mbf{A}_{\mathrm{d}}\mbf{T},\mbf{T}^\trans\mbf{B}_{\mathrm{d}},\mbf{C}_{\mathrm{d}}\mbf{T},\mbf{D}_{\mathrm{d}},\mbf{T}^\trans\mbf{P}_{\mathrm{d}}\mbf{T}\mright).
    \end{align}
    If \(\mbs{\theta}_{\mathrm{d}}=(\mbf{E}, \mbf{A}_{\mathrm{d}}, \mbf{B}_{\mathrm{d}}, \mbf{C}_{\mathrm{d}}, \mbf{D}_{\mathrm{d}}, \mbf{P}_{\mathrm{d}})\) satisfies \(\mbf{M}_{\eta,\mu}(\mbs{\theta}_{\mathrm{d}}) \succeq 0\), then
    \begin{align*}
        \diag\mleft(\mbf{T},\eye,\mbf{T}\mright)^{\trans} \mbf{M}_{\eta,\mu}(\mbs{\theta}_{\mathrm{d}}) \diag\mleft(\mbf{T},\eye,\mbf{T}\mright) = \mbf{M}_{\eta,\mu}(\mbs{\theta}_\mbf{T}) \succeq 0.
    \end{align*}

    This coordinate freedom can introduce arbitrary scaling and poor conditioning in the optimization variables. As such, for a fixed \(\eta \in (0, 1)\), \(\mu \geq 0\), and \(\epsilon \in (0, 1]\), define the closed and convex feasible set
    \begin{align}\label{eq:theta}
        \mbs{\Theta}_{\mathrm{d}} \triangleq \mleft\{\mbs{\theta}_{\mathrm{d}} \mid \mbf{M}_{\eta, \mu}(\mbs{\theta}_{\mathrm{d}}) \succeq 0,\;\mbf{P}_{\mathrm{d}} \succeq \epsilon\eye,\; \trace (\mbf{P}_{\mathrm{d}}) = n\mright\}.
    \end{align}
    The trace normalization removes the scaling \(\mbf{T} = \kappa\eye\), \(\kappa > 0\), while the lower bound on \(\mbf{P}_{\mathrm{d}}\) removes the strict inequality. Combined, these two constraints impose the upper bound \(\mbf{P}_{\mathrm{d}} \preceq \bar{p}\eye\), where \(\bar{p} \triangleq n - (n-1)\epsilon\).

\subsection{Sequential Convex Programming}\label{sec:scp}
    The passivity constraint \cref{eq:desc_lmi} is convex in \(\mbs{\theta}_{\mathrm{d}}\), whereas the least-squares cost \cref{eqn:least_squares} remains nonconvex. {Moreover, \cref{eq:desc_lmi} is not conservative relative to \cref{eq:desc_kyp}, since any realization satisfying \cref{eq:desc_kyp} can be lifted to a transfer-equivalent point satisfying \cref{eq:desc_lmi}, as shown in \Cref{sec:initialization}. Hence, only the cost is approximated, through the Gauss--Newton (GN) SCP method summarized in \Cref{alg:scp}.}

At each iteration, the transfer matrix is linearized about the current iterate, and the resulting convex least-squares surrogate is minimized over \(\mbs{\Theta}_{\mathrm{d}}\). Starting from a feasible initialization, Armijo backtracking preserves feasibility by convexity of \(\mbs{\Theta}_{\mathrm{d}}\), so every iterate is a passive model with margin \(\mu\) and the method can be terminated at any iteration.

    \begin{algorithm}[t]
    \caption{Affine Positive Real Constrained SCP}\label{alg:scp}
    \small
    \begin{algorithmic}[1]
        \Statex \textbf{Input:}
        \(\{(\omega_k,\mbf{Y}_k)\}_{k=1}^{N_f}\), \(n\), \(\mbs{\theta}_{\mathrm{d},0}\in\mbs{\Theta}_{\mathrm{d}}\), \(\tau>1\), \(\epsilon_{\mathrm{step}}>0\), \(0<\gamma_{\min}\leq\gamma_0\leq\gamma_{\max}\), \(c,\nu\in(0,1)\), and \(I_{\max}\)
        \For{\(i=0,\ldots,I_{\max}-1\)}
            \State Solve \cref{eq:subproblem} for \(\hat{\mbs{\theta}}_{\mathrm{d}, i}\); \(\mbf{d}_i \gets \hat{\mbs{\theta}}_{\mathrm{d}, i} - \mbs{\theta}_{\mathrm{d}, i}\)
            \If{\(\smash[b]{\norm{\mbf{d}_i}_{\mbf{P}_{\mathrm{d}, i}}} \leq \epsilon_{\mathrm{step}}\)} \textbf{break}%
            \EndIf
            \State \(\alpha_i\gets1\)
            \While{%
                \(J(\mbs{\theta}_{\mathrm{d}, i} + \alpha_i\mbf{d}_i) > J(\mbs{\theta}_{\mathrm{d}, i}) + c\alpha_i J'(\mbs{\theta}_{\mathrm{d}, i};\mbf{d}_i)\)
            }%
                \State \(\alpha_i\gets\nu\alpha_i\)
            \EndWhile
            \State \(\mbs{\theta}_{\mathrm{d}, i+1}\gets\mbs{\theta}_{\mathrm{d}, i}+\alpha_i\mbf{d}_i\)
            \If{\(\alpha_i=1\)}
                \(\gamma_{i+1}\gets\max\{\gamma_{\min},\gamma_i/\tau\}\)
            \Else\hspace{2pt}
                \(\gamma_{i+1}\gets\min\{\gamma_{\max},\tau\gamma_i\}\)
            \EndIf
        \EndFor
        \Statex \textbf{Output:}
        \(\mleft(\mbf{E}^{-1}\mbf{A}_{\mathrm{d}}, \mbf{E}^{-1}\mbf{B}_{\mathrm{d}}, \mbf{C}_{\mathrm{d}}, \mbf{D}_{\mathrm{d}}\mright)\) at the last iterate
    \end{algorithmic}
\end{algorithm}

    Consider the nonconvex constrained identification problem
    \begin{equation}\label{eq:problem}\tag{OP1}
        \min_{\mbs{\theta}_{\mathrm{d}} \in \mbs{\Theta}_{\mathrm{d}}} \quad J(\mbs{\theta}_{\mathrm{d}}) = \sum_{k=1}^{N_f} \norm{\mbf{Y}_k-\mbf{G}(e^{\j\omega_k})}_{\frob}^{2}.
    \end{equation}
    By \Cref{prop:desc_lmi}, every \(\mbs{\theta}_{\mathrm{d}} \in \mbs{\Theta}_{\mathrm{d}}\) defines a positive real system with margin \(\mu\). Moreover, \(e^{\j\omega_k}\mbf{E}-\mbf{A}_{\mathrm{d}}\) is nonsingular for every \(k\), so the objective \(J\) is well-defined and smooth on \(\mbs{\Theta}_{\mathrm{d}}\). Let \(\mbs{\theta}_{\mathrm{d}, i} \triangleq (\mbf{E}_i, \mbf{A}_{\mathrm{d},i}, \mbf{B}_{\mathrm{d},i}, \mbf{C}_{\mathrm{d},i}, \mbf{D}_{\mathrm{d},i}, \mbf{P}_{\mathrm{d},i}) \in \mbs{\Theta}_{\mathrm{d}}\) be the current iterate. A feasible GN step is obtained as follows.

    The first-order Taylor expansion of \( \mbf{G}(z_k)\) about \(\mbs{\theta}_{\mathrm{d}, i}\) is given by
    \begin{align}\label{eq:lin}
        \mbc{L}_k(\mbs{\theta}_{\mathrm{d}}) \triangleq \mbf{C}_{\mathrm{d}}\mbs{\Phi}_k + \mbs{\Gamma}_k\mbf{B}_{\mathrm{d}} + \mbf{D}_{\mathrm{d}} + \mbs{\Gamma}_k(\mbf{A}_{\mathrm{d}} - z_k\mbf{E})\mbs{\Phi}_k,
    \end{align}
    where \(z_k \triangleq e^{\j\omega_k}\), \(\mbs{\Psi}_k \triangleq (z_k\mbf{E}_i - \mbf{A}_{\mathrm{d},i})^{-1}\), \(\mbs{\Gamma}_k \triangleq \mbf{C}_{\mathrm{d},i}\mbs{\Psi}_k\), and \(\mbs{\Phi}_k \triangleq\mbs{\Psi}_k\mbf{B}_{\mathrm{d},i}\). Next, a full-step trial point is obtained from the semidefinite program (SDP)
    \begin{align}\label{eq:subproblem}
        \hspace{-6pt}
        \hat{\mbs{\theta}}_{\mathrm{d}, i} \triangleq \argmin_{\mbs{\theta}_{\mathrm{d}} \in \mbs{\Theta}_{\mathrm{d}}}\sum_{k=1}^{N_f} \norm{\mbf{Y}_k - \mbc{L}_k(\mbs{\theta}_{\mathrm{d}})}_{\frob}^2 + \gamma_{i} \norm{\Delta \mbs{\theta}_{\mathrm{d}, i}}_{\mbf{P}_{\mathrm{d},i}}^2\hspace{-1pt},
    \end{align}
    where \(\Delta\mbs{\theta}_{\mathrm{d}, i} \triangleq \mbs{\theta}_{\mathrm{d}} - \mbs{\theta}_{\mathrm{d}, i}\), \(\gamma_{i} > 0\) is a Levenberg--Marquardt-type damping parameter, and
    \begin{align*}
        \norm{\Delta \mbs{\theta}_{\mathrm{d}, i}}_{\mbf{P}_{\mathrm{d},i}}^2
        &\triangleq
        \norm{\mbf{L}_i\Delta\mbf{E}_i\mbf{L}_i}_\frob^2 + \norm{\mbf{L}_i\Delta\mbf{A}_{\mathrm{d}, i}\mbf{L}_i}_\frob^2  \\%
        &\qquad\hspace{-3pt}+ \norm{\mbf{L}_i\Delta\mbf{B}_{\mathrm{d}, i}}_\frob^2 + \norm{\Delta\mbf{C}_{\mathrm{d},i}\mbf{L}_i}_\frob^2 + \norm{\Delta\mbf{D}_{\mathrm{d},i}}_\frob^2 \\%
        &\qquad\hspace{-3pt}+\norm{\mbf{L}_i\Delta\mbf{P}_{\mathrm{d},i}\mbf{L}_i}_\frob^2,
    \end{align*}
    with \(\mbf{L}_i \triangleq \mbf{P}_{\mathrm{d},i}^{-1/2}\). The proximal weighting norm, \(\norm{\cdot}_{\mbf{P}_{\mathrm{d},i}}\), is motivated by the transformation-invariant metric on the positive-definite cone introduced in \cite[Thm.~3.1]{ohara1994differential}. Importantly, this metric is invariant when the base point \(\mbs{\theta}_{\mathrm{d}, i}\) and the trial point \(\mbs{\theta}_{\mathrm{d}}\) are both transformed by the same congruence \cref{eqn:congurance}. Moreover, since \(\epsilon\eye \preceq \mbf{P}_{\mathrm{d},i} \preceq \bar{p}\eye\), it follows that \(\norm{\Delta \mbs{\theta}_{\mathrm{d}, i}}_{\mbf{P}_{\mathrm{d},i}}^2 \geq \bar{p}^{-2}\norm{\Delta \mbs{\theta}_{\mathrm{d}, i}}_{\frob}^2\). Therefore, the objective in \cref{eq:subproblem} is strongly convex and coercive. Consequently, \cref{eq:subproblem} admits a unique minimizer \(\hat{\mbs{\theta}}_{\mathrm{d}, i}\) for every \(\gamma_i > 0\).

    For \(\mbf{d}_i \triangleq\hat{\mbs{\theta}}_{\mathrm{d}, i}-\mbs{\theta}_{\mathrm{d}, i}\neq0\), comparison of the subproblem objective at \(\hat{\mbs{\theta}}_{\mathrm{d}, i}\) and the feasible point \(\mbs{\theta}_{\mathrm{d}, i}\), with directional derivatives taken over the real decision variables, gives
    \begin{align}\label{eq:descent}
        J'(\mbs{\theta}_{\mathrm{d}, i};\mbf{d}_i)%
        &\mathrel{{=}}{-2\sum_{k=1}^{N_f}\mathrm{Re}\,\trace\mleft(\mleft(\mbf{Y}_k-\mbf{G}(z_k)\mright)^{\herm}\mbc{L}_k(\mbf{d}_i)\mright)}\nonumber\\%
        &\leq - \sum_{k=1}^{N_f}\norm{\mbc{L}_k(\mbf{d}_i)}_{\frob}^{2} -\gamma_{i}\norm{\mbf{d}_i}_{\mbf{P}_{\mathrm{d},i}}^{2} < 0.
    \end{align}

    {Next, let \(\hat{J}_{i}(\mbs{\theta}_{\mathrm{d}})\) denote the objective in \cref{eq:subproblem}. Since \(\mbc{L}_k\) in \cref{eq:lin} is linear with \(\mbc{L}_k(\mbs{\theta}_{\mathrm{d},i}) = \mbf{G}(z_k)\) and the proximal term has zero gradient at \(\mbs{\theta}_{\mathrm{d},i}\), it follows that \(\hat{J}_{i}'(\mbs{\theta}_{\mathrm{d},i};\Delta\mbs{\theta}_{\mathrm{d},i}) = J'(\mbs{\theta}_{\mathrm{d},i};\Delta\mbs{\theta}_{\mathrm{d},i})\) for all \(\mbs{\theta}_{\mathrm{d}}\in\mbs{\Theta}_{\mathrm{d}}\). Since \cref{eq:subproblem} is convex with a unique minimizer, \(\mbf{d}_i = 0\) if and only if \(J'(\mbs{\theta}_{\mathrm{d},i};\Delta\mbs{\theta}_{\mathrm{d},i}) \geq 0\) for all \(\mbs{\theta}_{\mathrm{d}}\in\mbs{\Theta}_{\mathrm{d}}\), that is, \(\mbs{\theta}_{\mathrm{d},i}\) is first-order stationary for \cref{eq:problem}.} Since \(\mbs{\Theta}_{\mathrm{d}}\) is convex, a backtracking line search ensures that the next iterate \(\mbs{\theta}_{\mathrm{d},i+1} = \mbs{\theta}_{\mathrm{d}, i} + \alpha_i\mbf{d}_i\) remains feasible and the smoothness of \(J\) guarantees finite Armijo backtracking. {Moreover, every limit point of the iterates is first-order stationary for \cref{eq:problem} \cite[Prop.~2.2.1]{bertsekas1999nonlinear}.}

    \subsection{Feasible Initialization}\label{sec:initialization}
    An initial unconstrained estimate is obtained from FSID. {To ensure} \(\rho(\mbf{A}_0)<r\), with \(r\) given by \cref{eq:radius}, {any pole \(p\) with \(\lvert p \rvert > \bar{r}\) is replaced by \(\bar{r}p/\lvert p \rvert\), where \(\bar{r} < r\).} Let \((\mbf{A}_0,\mbf{B}_0)\) denote the resulting dynamics. For a fixed \(\eta \in (0, 1)\) and \(\mu \geq 0\), a positive real system with margin \(\mu\) is then obtained by solving the following convex SDP
    \begin{equation}\label{alg:warm-start}\tag{OP2}
        \hspace{-10pt}
        \begin{aligned}
            \min_{\mbf{C}, \mbf{D}, \mbf{P}} &\quad J(\mbs{\theta})\\
            \text{s.t.} &\quad \mbf{M}_{\mathrm{KYP}}(\mbs{\theta}) \succeq \diag\mleft((1-r^2)\mbf{P},\mu\eye\mright),\\
            &\quad \mbf{P} \succeq \epsilon_{\mbf{P}}\eye,
        \end{aligned}
    \end{equation}
    where \(\mbs{\theta}=(\mbf{A}_0, \mbf{B}_0, \mbf{C}, \mbf{D}, \mbf{P})\), \(\epsilon_{\mbf{P}} > 0\).
    \begin{remark}[Feasibility of \cref{alg:warm-start}]
        Since \(\rho(\mbf{A}_0)<r\) by construction, there exists \(\mbf{P} \succeq \epsilon_{\mbf{P}}\eye\) satisfying \(\mbf{A}_0^\trans\mbf{P}\mbf{A}_0 \prec r^2\mbf{P}\). Finally, choosing \(\mbf{C} = \mbf{B}_0^\trans\mbf{P}\mbf{A}_0\) and \(\mbf{D} = \tfrac{1}{2}(\mbf{B}_0^\trans\mbf{P}\mbf{B}_0 + \mu\eye)\) provides a feasible point.
    \end{remark}

    To obtain a feasible initialization for \Cref{alg:scp}, let \((\mbf{C}_0,\mbf{D}_0,\mbf{P}_0)\) solve \cref{alg:warm-start} and define the descriptor lift
    \begin{align*}
        \bar{\mbs{\theta}}_{\mathrm{d},0} \triangleq \mleft(
            r^2\mbf{P}_0, r^2\mbf{P}_0\mbf{A}_0, r^2\mbf{P}_0\mbf{B}_0, \mbf{C}_0, \mbf{D}_0, \tfrac{r^2}{1+\eta}\mbf{P}_0\mright).
    \end{align*}
    It can be verified that \(\bar{\mbs{\theta}}_{\mathrm{d},0}\) satisfies \(\mbf{M}_{\eta, \mu}(\bar{\mbs{\theta}}_{\mathrm{d}, 0}) \succeq 0\) and realizes the same transfer matrix as \((\mbf{A}_0,\mbf{B}_0,\mbf{C}_0,\mbf{D}_0)\). {The same lift applies to any realization satisfying \cref{eq:desc_kyp}.} To satisfy the trace normalization in \cref{eq:theta}, the congruence transformation \cref{eqn:congurance} with \(\mbf{T}=\tfrac{\sqrt{1+\eta}}{r}\mbf{P}_0^{-1/2}\) is applied to \(\bar{\mbs{\theta}}_{\mathrm{d},0}\). The resulting point \(\mbs{\theta}_{\mathrm{d},0}\in\mbs{\Theta}_{\mathrm d}\) satisfies \(\mbf{E}_0=(1+\eta)\eye\) and \(\mbf{P}_{\mathrm{d},0}=\eye\) and is used to initialize \Cref{alg:scp}.

     \section{Discussion} \label{sec:discussion}
    The general formulation in \Cref{sec:scp} treats both \(\mbf{E}\) and \(\mbf{P}_{\mathrm{d}}\) as decision variables, allowing for additional affine constraints to be imposed in prescribed descriptor coordinates. For example, constraining \(\mbf{E}\) to be block diagonal ensures that compatible zero blocks in \(\mbf{A}_{\mathrm{d}}\) and \(\mbf{B}_{\mathrm{d}}\) are preserved in \(\mbf{A}=\mbf{E}^{-1}\mbf{A}_{\mathrm{d}}\) and \(\mbf{B}=\mbf{E}^{-1}\mbf{B}_{\mathrm{d}}\), respectively. In the \emph{unstructured} setting, such constraints are not imposed, and the state coordinates are arbitrary. Therefore, the congruence freedom in \cref{eqn:congurance} enlarges the search space without enlarging the set of attainable transfer matrices. {Hence, the general formulation is needed only for structured identification.}

\subsection{Upper Hessenberg Form}\label{subsec:unstructured}
    In the unstructured setting, the descriptor matrix \(\mbf{E}\) and the certificate \(\mbf{P}_{\mathrm{d}}\) can be normalized to reduce the number of decision variables without changing the set of attainable transfer matrices. Specifically, consider the normalization
    \begin{align}\label{eqn:unstructured_normalization}
        \mbf{E} &= \eye, & \mbf{P}_{\mathrm{d}} &= \frac{1}{1+\eta}\eye.
    \end{align}
    By \Cref{prop:desc_lmi}, any feasible descriptor realization yields a transfer-equivalent state-space realization \((\mbf{A}, \mbf{B}, \mbf{C}, \mbf{D})\) and a certificate \(\mbf{P}\succ0\) satisfying the shifted KYP inequality \cref{eq:desc_kyp}. Let \(\mbf{T} = r^{-1}\mbf{P}^{-1/2}\) with \(r\) given by \cref{eq:radius}. The state-coordinate transformation \(\bar{\mbf{A}} = \mbf{T}^{-1}\mbf{A}\mbf{T}\), \(\bar{\mbf{B}} = \mbf{T}^{-1}\mbf{B}\), \(\bar{\mbf{C}} = \mbf{C}\mbf{T}\), and \(\bar{\mbf{D}} = \mbf{D}\) preserves the transfer matrix. The transformed certificate is \(\bar{\mbf{P}} = \mbf{T}^{\trans}\mbf{P}\mbf{T} = r^{-2}\eye\). Consequently, the transformed shifted KYP inequality is equivalent, by a Schur complement, to
    \begin{align*}
        \begin{bmatrix}
            \eye          & \bar{\mbf{C}}^\trans                           & \bar{\mbf{A}}^\trans \\
            \bar{\mbf{C}} & \bar{\mbf{D}} + \bar{\mbf{D}}^\trans - \mu\eye & \bar{\mbf{B}}^\trans \\
            \bar{\mbf{A}} & \bar{\mbf{B}}                                  & r^2\eye
        \end{bmatrix}
        \succeq 0.
    \end{align*}
    This is equivalent to the descriptor LMI, \(\mbf{M}_{\eta, \mu}(\mbs{\theta}_{\mathrm{d}}) \succeq 0\), with \(\mbs{\theta}_{\mathrm{d}} = (\eye, \bar{\mbf{A}}, \bar{\mbf{B}}, \bar{\mbf{C}}, \bar{\mbf{D}}, \tfrac{1}{1+\eta}\eye)\). Hence, every transfer matrix admitted by \cref{eq:desc_lmi} is also attained under the normalization \cref{eqn:unstructured_normalization}.

    Under the normalization \cref{eqn:unstructured_normalization}, orthogonal coordinate freedom remains in the form of \(\mbf{U}\) satisfying \(\mbf{U}^{\trans}\mbf{U} = \eye\). Notably, every real square matrix admits an orthogonal similarity transformation to upper Hessenberg form \cite[Sec.~7.4.3]{hessenberg}. Therefore, an upper Hessenberg structure can be imposed on \(\mbf{A}_{\mathrm{d}}\) by requiring \([\mbf{A}_{\mathrm{d}}]_{ij}=0\) for all \(i>j+1\), where \([\mbf{A}_{\mathrm{d}}]_{ij}\) denotes the \((i,j)\) entry of \(\mbf{A}_{\mathrm{d}}\). The linear equality constraints on \(\mbf{A}_{\mathrm{d}}\) eliminate \((n-1)(n-2)/2\) scalar decision variables from the optimization problem.

    For the unstructured formulation, the normalization constraints in \cref{eq:theta} are replaced by \cref{eqn:unstructured_normalization}, and the Hessenberg structure is imposed by defining
    \begin{equation}\label{eq:theta_H}
        \hspace{-5pt}
        \mbs{\Theta}_{\mathrm{H}} \triangleq
        \left\{
        \mbs{\theta}_{\mathrm{d}} \;\middle|\;
        \begin{aligned}
            &\mbf{M}_{\eta,\mu}(\mbs{\theta}_{\mathrm{d}}) \succeq 0, \mbf{E} = \eye, \mbf{P}_{\mathrm{d}} = \frac{1}{1+\eta}\eye,\\
            &[\mbf{A}_{\mathrm{d}}]_{ij} = 0, \text{ for } i > j+1
        \end{aligned}
        \right\}.
    \end{equation}
    The initial point constructed in \Cref{sec:initialization} is mapped into \(\mbs{\Theta}_{\mathrm{H}}\) by applying the descriptor congruence transformation \cref{eqn:congurance} with \(\mbf{T} = (1+\eta)^{-1/2}\mbf{U}\), where \(\mbf{U}^{\trans}\mbf{U} = \eye\) and \(\mbf{U}^{\trans}\mbf{A}_{\mathrm{d},0}\mbf{U}\) is upper Hessenberg. The SCP method in \Cref{alg:scp} is then applied over \(\mbs{\Theta}_{\mathrm{H}}\), with \(\mbf{E}\) and \(\mbf{P}_{\mathrm{d}}\) removed from the decision variables and the Hessenberg equality constraints imposed in each subproblem.

    \section{Numerical Results}\label{sec:simulation}
    {%
    The proposed method is evaluated using noisy frequency-response data from randomly generated positive real plants. For each plant, let \(\mbf{F} = \mbf{U}\mbs{\Sigma}\mbf{W}^{\trans}\) be a singular value decomposition of \(\mbf{F} \in \mathbb{R}^{6 \times 6}\) with independent \(\mathcal{N}(0, 1)\) entries, and let \(\xi_\ell \sim \mathcal{N}(0, 1)\), for \(\ell = 1, \ldots, 6\), be independent. Define \(\mbf{Z} \triangleq \mbf{U}\diag\mleft(\lambda_1, \ldots, \lambda_6\mright)\mbf{U}^{\trans}\), where \(\lambda_\ell \triangleq e^{\xi_\ell}\). For a pole radius \(\rho_\star \in (0, 1)\), consider the realization
    \begin{equation}\label{eq:sim_plant}
        \begin{gathered}
            \mbf{S} \triangleq
            \begin{bmatrix}
                \mbf{0} & \mbf{Z} \\
                -\mbf{Z} & \mbf{0}
            \end{bmatrix}, \quad
            \mbf{A}_\star \triangleq \rho_\star(\eye + \mbf{S})(\eye - \mbf{S})^{-1}, \\
            \mbf{B}_\star \triangleq \begin{bmatrix} \eye & \mbf{0} \end{bmatrix}^{\trans}, \quad
            \mbf{C}_\star \triangleq 2\mbf{B}_\star^{\trans}\mbf{A}_\star, \quad \mbf{D}_\star \triangleq \eye,
        \end{gathered}
    \end{equation}
    where \(\mbf{B}_\star \in \mathbb{R}^{12 \times 3}\). The corresponding transfer matrix is \(\mbf{G}_\star(z) = \mbf{B}_\star^{\trans}(z\eye + \mbf{A}_\star)(z\eye - \mbf{A}_\star)^{-1}\mbf{B}_\star\), with \(n_\star = 12\).

    Since \(\mbf{S}\) is skew-symmetric, \((\eye + \mbf{S})(\eye - \mbf{S})^{-1}\) is orthogonal. Therefore, \(\mbf{A}_\star^{\trans}\mbf{A}_\star = \rho_\star^2\eye\), and the poles of \(\mbf{G}_\star\) are \(\rho_\star e^{\pm 2\j\arctan\lambda_\ell}\), for \(\ell = 1, \ldots, 6\). The pole angles vary between plants, while the pole radius is fixed. Moreover, the poles are distinct and \cref{eq:sim_plant} is minimal almost surely. It can be shown that \(\mbf{G}_\star(z) + \mbf{G}_\star(z)^{\herm} \succeq \tfrac{2(\lvert z \rvert - \rho_\star)}{\lvert z \rvert + \rho_\star}\eye\) for all \(\lvert z \rvert > \rho_\star\). Hence, \(\mbf{G}_\star\) is positive real with a margin of at least \(2(1 - \rho_\star)/(1 + \rho_\star)\).

    To make the identification problem more challenging, the benchmark uses \(\rho_\star = 0.9995\), for which the margin of \(\mbf{G}_\star\) is at least \(5.0 \times 10^{-4}\). The training grid contains \(N_f = 512\) uniformly spaced frequencies \(\omega_k = (k-1)\pi/(N_f - 1)\).

    For a signal-to-noise ratio (SNR) of \(s\,\mathrm{dB}\), the measurements are \(\mbf{Y}_k \triangleq \mbf{G}_\star(e^{\j\omega_k}) + \sigma_s\mbf{V}_k\), where \(\sigma_s > 0\) is chosen such that \(\sum_{k=1}^{N_f}\norm{\sigma_s\mbf{V}_k}_{\frob}^{2} = 10^{-s/10}\sum_{k=1}^{N_f}\norm{\mbf{G}_\star(e^{\j\omega_k})}_{\frob}^{2}\). The entries of \(\mbf{V}_k\) are independent of each other and across frequencies. For \(0 < \omega_k < \pi\), their real and imaginary parts are independent and distributed as \(\mathcal{N}(0, \tfrac{1}{2})\), and for \(\omega_k \in \{0, \pi\}\), they are real and distributed as \(\mathcal{N}(0, 1)\). This model is standard for frequency-response data~\cite{mckelvey1996subspace} and holds for steady-state measurements over complete periods with white output noise when each experiment excites one input with the same flat-spectrum multisine. Ten randomly generated plants are evaluated using five common noise realizations for each \(s \in \{10, 20, 30\}\), resulting in \(50\) plant--noise combinations per SNR\@. The same plants and noise realizations are used at every SNR, and the same data are used across all methods.

    Let \(\widehat{\mbf{G}}\) denote the transfer matrix of an identified model of order \(n \leq n_\star\), and let \(\widehat{\mbf{G}}_k \triangleq \widehat{\mbf{G}}(e^{\j\bar{\omega}_k})\) and \(\mbf{G}_{\star,k} \triangleq \mbf{G}_\star(e^{\j\bar{\omega}_k})\) on a uniform validation grid \(\bar{\omega}_k = (k-1)\pi/(\bar{N}_f - 1)\) with \(\bar{N}_f = 2^{20}\) points. The relative validation error is \(E_2 \triangleq\nobreak \bigl(\sum_{k=1}^{\bar{N}_f}\lVert\widehat{\mbf{G}}_k - \mbf{G}_{\star,k}\rVert_{\frob}^{2} \big/ \sum_{k=1}^{\bar{N}_f}\lVert\mbf{G}_{\star,k}\rVert_{\frob}^{2}\bigr)^{1/2}\).
    }

    As an unconstrained baseline, the implementation of FSID in \cite{MCKELVEY2021403} is used. The unconstrained GN method starts from the FSID estimate and jointly optimizes \((\mbf{A}, \mbf{B}, \mbf{C}, \mbf{D})\) to minimize \cref{eqn:least_squares}. The realization is rebalanced after each accepted step to improve numerical conditioning.

    Two post-processing methods, FSID-PR and GN-PR, are considered. By solving the SDP in \cref{alg:warm-start}, these methods enforce positive realness with margin \(\mu = 10^{-6}\) on the FSID and final GN estimates, respectively. Both methods fix \((\mbf{A}, \mbf{B})\) and refit \((\mbf{C}, \mbf{D})\). Hence, they cannot adjust the pole locations.

    The pH-Flex method is the flexible port-Hamiltonian parametrization in \cite{schwerdtner2021}, adapted to minimize the frequency-domain cost in \cref{eqn:least_squares} for discrete-time systems using the passivity-preserving bilinear transformation. It uses \(10\)  Broyden--Fletcher--Goldfarb--Shanno (BFGS) restarts capped at \(2000\) iterations each.

    The proposed \Cref{alg:scp} is initialized following~\Cref{sec:initialization} and is referred to as PR-SCP\@. The unstructured formulation in~\Cref{subsec:unstructured} is denoted PR-SCP-H, where \(\mbf{A}_{\mathrm{d}}\) is constrained to be in upper Hessenberg form. Both SCP algorithms use the same parameters, \(\eta=10^{-6}\), \(\mu = 10^{-6}\), {\(\epsilon = 0.1\), \(\epsilon_{\mbf{P}} = 10^{-6}\), \(\bar{r} = 0.99998\), \(\gamma_0 = 10^{-3}\),} \(\mleft(\gamma_{\min}, \gamma_{\max}\mright) = \mleft(10^{-4}, 10^{6}\mright)\), {\(\tau = 5\),} and \((c, \nu ) = (10^{-4}, 0.5)\). All GN-type methods use the same stopping tolerance \(\epsilon_\mathrm{step} = {10^{-9}}\) and maximum number of iterations \(I_{\mathrm{max}} = {100}\). Finally, the SDPs are solved with \texttt{MOSEK} \cite{MOSEK} through \texttt{CVXPY} \cite{CVXPY}.

\begin{figure}[t]
        \centering
        \includegraphics{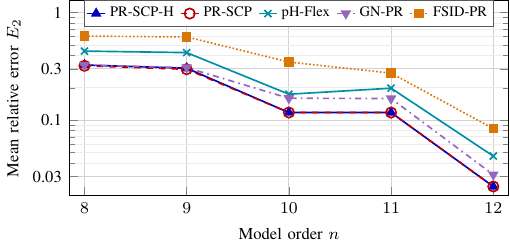}
        \caption{Mean relative validation error \(E_2\) of the passivity-enforcing methods versus model order \(n\) at \(20\,\mathrm{dB}\), on a logarithmic vertical axis.}
        \label{fig:sim_fir_order}
    \end{figure}

    {\Cref{tab:error_results} shows that PR-SCP and PR-SCP-H attain the lowest mean \(E_2\) at every SNR\@. The unconstrained GN method attains comparable errors, but it did not produce a passive model in any tested case. PR-SCP-H matches the errors of PR-SCP in a fraction of the computation time, since the normalization and Hessenberg restriction reduce the number of decision variables without changing the attainable transfer matrices. For reduced-order modeling at \(20\,\mathrm{dB}\), \Cref{fig:sim_fir_order} shows that the proposed methods also attain the lowest mean \(E_2\).}

    \begin{table}[t]
    \vspace{7pt}
    \centering
    \caption{Full-order (\(n = 12\)) validation error \(E_2\) (mean \(\pm\) standard deviation) and mean identification time \(t\).}
    \scriptsize
    \setlength{\tabcolsep}{2.5pt}
    \label{tab:error_results}
    \begin{tabular*}{\linewidth}{@{\extracolsep{\fill}}lcccc@{}}
        \hline\hline\addlinespace[1pt]
        Method   & 10 dB             & 20 dB             & 30 dB             & \(t\) [s]\\%
        \hline\addlinespace[2pt]
        FSID     & $0.299 \pm 0.096$ & $0.079 \pm 0.029$ & $0.022 \pm 0.006$ & $0.08$\\
        FSID-PR  & $0.313 \pm 0.099$ & $0.083 \pm 0.031$ & $0.026 \pm 0.010$ & $0.64$\\
        GN       & $0.080 \pm 0.028$ & $0.025 \pm 0.007$ & $0.008 \pm 0.002$ & $69.6$\\
        GN-PR    & $0.089 \pm 0.028$ & $0.031 \pm 0.011$ & $0.011 \pm 0.003$ & $70.3$\\
        pH-Flex  & $0.099 \pm 0.047$ & $0.046 \pm 0.014$ & $0.010 \pm 0.003$ & $289.3$\\
        PR-SCP   & $0.072 \pm 0.022$ & $0.024 \pm 0.007$ & $0.008 \pm 0.002$ & $302.0$\\
        PR-SCP-H & $0.072 \pm 0.022$ & $0.024 \pm 0.007$ & $0.008 \pm 0.002$ & $40.1$\\
        \hline\hline
    \end{tabular*}
\end{table}

    \section{Closing Remarks} \label{sec:closing_remarks}
    The identification of passive discrete-time LTI systems from frequency-response data is considered in this paper. An affine descriptor LMI enforces a prescribed passivity margin and spectral-radius bound. A feasible Gauss--Newton SCP method jointly refines the realization matrices. For unstructured identification, normalization and an upper Hessenberg restriction reduce the number of decision variables without changing the attainable transfer matrices. Numerical results on randomized passive plants show that the proposed methods attain the lowest mean validation error among the passivity-enforcing methods considered, at both full and reduced model orders. {Future work will apply PR-SCP to structured identification, such as networks with a prescribed topology.}

    \printbibliography
    \onecolumn\section{Additional Simulation Results}
\subsection{Supplementary Results for the Benchmark in \texorpdfstring{\Cref{sec:simulation}}{Section \ref*{sec:simulation}}}\label{subsec:benchmark_extra}
    This subsection continues the benchmark in \Cref{sec:simulation} and reports additional metrics for the same experiments. The following tables use the same full-order fits as \Cref{tab:error_results}, with \(50\) plant--noise combinations per SNR\@. The relative training error is measured against the noisy frequency-response data:
    \begin{equation*}
    E_{\mathrm{tr}}\triangleq
    \left(\frac{\sum_{k=1}^{N_f}\|\widehat{\mbf{G}}(e^{\j\omega_k})-\mbf{Y}_k\|_{\frob}^{2}}
    {\sum_{k=1}^{N_f}\|\mbf{Y}_k\|_{\frob}^{2}}\right)^{1/2}.
    \end{equation*}
    \begin{table}[H]
        \centering
        \caption{Full-order (\(n=12\)) relative training error \(E_{\mathrm{tr}}\) (mean \(\pm\) sample standard deviation) and mean identification time \(t\), averaged over the three SNRs.}
        \label{tab:training_results}
        \footnotesize
        \setlength{\tabcolsep}{5pt}
        \begin{tabular*}{\linewidth}{@{\extracolsep{\fill}}lcccc@{}}
        \hline\hline\addlinespace[1pt]
        Method & 10 dB & 20 dB & 30 dB & \(t\) [s]\\
        \hline\addlinespace[2pt]
        FSID & $0.30547 \pm 0.00358$ & $0.09981 \pm 0.00031$ & $0.03170 \pm 0.00021$ & $0.08$\\
        FSID-PR & $0.31070 \pm 0.00791$ & $0.10091 \pm 0.00340$ & $0.03179 \pm 0.00031$ & $0.64$\\
        GN & $0.30021 \pm 0.00113$ & $0.09913 \pm 0.00017$ & $0.03147 \pm 0.00002$ & $69.6$\\
        GN-PR & $0.30065 \pm 0.00116$ & $0.09921 \pm 0.00019$ & $0.03148 \pm 0.00002$ & $70.3$\\
        pH-Flex & $0.30649 \pm 0.00687$ & $0.09970 \pm 0.00042$ & $0.03149 \pm 0.00003$ & $289.3$\\
        PR-SCP & $0.30044 \pm 0.00111$ & $0.09918 \pm 0.00017$ & $0.03148 \pm 0.00002$ & $302.0$\\
        PR-SCP-H & $0.30044 \pm 0.00111$ & $0.09918 \pm 0.00017$ & $0.03148 \pm 0.00002$ & $40.1$\\
        \hline\hline
        \end{tabular*}
    \end{table}
    A certified margin is computed for each fixed identified realization by solving an SDP that maximizes \(\mu\) over the storage certificate subject to the KYP inequality with margin \(\mu\). The mean margin \(\bar{\mu}\) is then obtained by averaging these certified margins over the \(50\) runs at each SNR\@.
    \begin{table}[H]
        \centering
        \caption{Full-order (\(n=12\)) mean passivity margin, \(\bar{\mu}\).}
        \label{tab:margin_results}
        \footnotesize
        \setlength{\tabcolsep}{5pt}
        \begin{tabular*}{\linewidth}{@{\extracolsep{\fill}}lccc@{}}
        \hline\hline\addlinespace[1pt]
        Method & 10 dB & 20 dB & 30 dB\\
        \hline\addlinespace[2pt]
        FSID & --- & --- & ---\\
        FSID-PR & $1.924\times10^{-6}$ & $1.105\times10^{-6}$ & $1.047\times10^{-6}$\\
        GN & --- & --- & ---\\
        GN-PR & $2.374\times10^{-6}$ & $1.168\times10^{-6}$ & $1.034\times10^{-6}$\\
        pH-Flex & $6.610\times10^{-4}$ & $3.346\times10^{-5}$ & $1.818\times10^{-5}$\\
        PR-SCP & $2.120\times10^{-4}$ & $5.477\times10^{-5}$ & $1.438\times10^{-5}$\\
        PR-SCP-H & $1.922\times10^{-4}$ & $4.981\times10^{-5}$ & $1.262\times10^{-5}$\\
        \hline\hline
        \end{tabular*}
    \end{table}
\subsection{FIR-Based Benchmark}\label{subsec:fir_benchmark}
    This subsection reports the benchmark used in the initial submission of this work. It considers randomly generated positive real plants with a prescribed passivity margin and pole location. Fix \(m > 1\) and \(N \in \mathbb{Z}_{>0}\), and let \(\Omega_\ell \triangleq \pi \ell / N\), for \(\ell = 0, \ldots, 2N - 1\), denote the \(2N\)-point discrete Fourier transform (DFT) grid. Let \(\mbf{U}_\ell \in \mathbb{C}^{m \times m}\) be the unitary factor in a QR factorization of an \(m \times m\) complex Gaussian matrix for \(0 < \ell < N\), and of an \(m \times m\) real Gaussian matrix for \(\ell \in \{0, N\}\). Define the Hermitian samples
    \begin{align}\label{eq:fir_samples}
        \mbf{Z}_\ell \triangleq \mu_\star\eye + m(1-\mu_\star)\mbf{U}_\ell \diag\mleft(0,w_{\ell,2},\ldots,w_{\ell,m}\mright) \mbf{U}_\ell^{\herm},
    \end{align}
    where \(\mu_\star\in(0,1)\) is the desired passivity margin and \((w_{\ell,2},\ldots,w_{\ell,m})\) is drawn uniformly from the set of nonnegative vectors whose entries sum to one. Set \(\mbf{Z}_{2N-\ell}=\mbf{Z}_\ell^\trans\) for \(\ell=1,\ldots,N-1\). Using this construction, every sample is Hermitian with minimum eigenvalue \(\mu_\star\) and \(\trace(\mbf{Z}_\ell)=m\). Moreover, the samples at \(\ell \in \{0, N\}\) are real, so the sequence is conjugate symmetric and its inverse DFT, \(\mbf{R}_t = \tfrac{1}{2N} \sum_{\ell=0}^{2N-1} \mbf{Z}_\ell e^{\j \Omega_\ell t}\), is real. Define the finite impulse response (FIR) transfer matrix
    \begin{align*}
        \mbf{G}_0(z) = \frac{1}{2} \mbf{R}_0 + \sum_{t=1}^{2N - 1} \frac{2N - t}{2N} \mbf{R}_t z^{-t}.
    \end{align*}
    On the unit circle, it can be verified that
    \begin{align}\label{eq:fir_fejer}
        \mbf{G}_0(e^{\j\omega}) + \mbf{G}_0(e^{\j\omega})^{\herm}
        = \frac{1}{2N} \sum_{\ell=0}^{2N - 1} K\mleft(\omega - \Omega_\ell\mright) \mbf{Z}_\ell,
    \end{align}
    where \(K(\omega) = \sin^2\mleft(N\omega\mright) / \mleft(2N \sin^2\mleft(\omega/2\mright)\mright)\) is the Fej\'er kernel, with \(K(0)=2N\). The Hermitian sum in \cref{eq:fir_fejer} is therefore a convex combination of the samples at every frequency and interpolates the samples on the grid. The minimum eigenvalue of the Hermitian sum over frequency is \(\mu_\star\). Therefore, the FIR transfer matrix \(\mbf{G}_0\) is positive real with margin exactly \(\mu_\star\).

    Since an FIR model has all of its poles at the origin, nontrivial dynamics are introduced through the all-pass substitution
    \begin{align}\label{eq:fir_warp}
        \mbf{G}_{\star}(z) = \mbf{G}_0\mleft(\frac{z - a}{1 - az}\mright), \qquad a \in (-1, 1).
    \end{align}
    The transformation in \cref{eq:fir_warp} is a bijection of the unit circle onto itself. Therefore, \(\mbf{G}_{\star}\) has the same passivity margin, \(\mu_\star\), as \(\mbf{G}_0\). The resulting system has a real state-space realization of order \(n_\star = m\mleft( 2N - 1 \mright)\) with all poles located at \(a\).

    The benchmark uses \(m=3\), \(N=4\), \(\mu_{\star}=10^{-4}\), and \(a=0.98\), which gives \(n_{\star}=21\). Fifty randomly generated plants are evaluated using two common noise seeds, resulting in \(100\) plant--noise combinations. For each combination, the same data are used across all methods and model orders. The training grid contains \(N_f = 128\) points, with \(\omega_1 = 0\) and the remaining \(N_f-1\) frequencies geometrically spaced over \([10^{-3},\pi]\). For \(\mbf{Y}_k \triangleq \mbf{G}_{\star}(e^{\j\omega_k})+\mbf{W}_k\), the additive Gaussian noise is real at \(\omega_k = 0\) and \(\omega_k = \pi\), and is scaled to a signal-to-noise ratio of \(20\,\mathrm{dB}\) over the training grid. Validation uses the noise-free response at \(\bar{N}_f=4096\) geometrically spaced frequencies, and \(E_2\) is defined as in \Cref{sec:simulation}. Model orders \(n=10,\ldots,21\) are evaluated, covering reduced-order approximations for \(n<21\) and the full-order identification case of \(n=n_\star=21\).

    The compared methods and their parameters are as described in \Cref{sec:simulation}, except that all GN-type methods use the stopping tolerance \(\epsilon_\mathrm{step} = 10^{-6}\) and maximum number of iterations \(I_{\mathrm{max}} = 50\). The mean margin \(\bar{\mu}\) is computed as in \Cref{subsec:benchmark_extra}, averaged over all \(100\) runs.

\begin{figure}[H]
        \centering
        \includegraphics{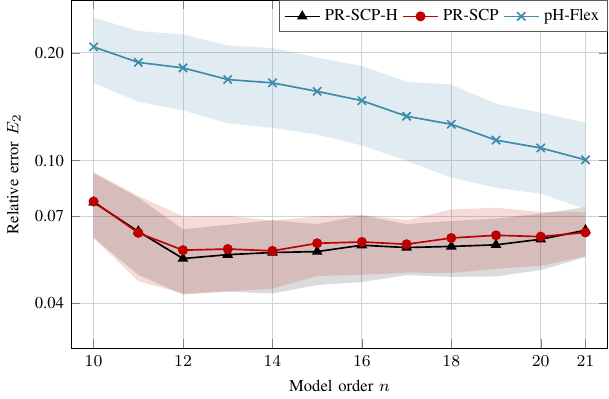}
        \caption{FIR-based benchmark: relative validation error \(E_2\) versus model order \(n\), on a logarithmic vertical axis. Curves show means over the \(100\) plant--noise combinations and shaded regions indicate \(\pm\) one standard deviation.}
        \label{fig:fir_order}
    \end{figure}

    \Cref{tab:fir_n21} shows that the proposed methods, PR-SCP and PR-SCP-H, achieve the two lowest mean \(E_2\) values at full order while providing certified passive models. PR-SCP-H substantially reduces the mean computation time while achieving similar validation errors to PR-SCP\@. Neither FSID nor GN produces a passive model in any tested case. In the reduced-order case, pH-Flex degrades more substantially than the proposed methods, as shown in \Cref{fig:fir_order}.

    \begin{table}[H]
        \centering
        \caption{FIR-based benchmark: full-order (\(n=21\)) results for fifty plants and two noise seeds. \(E_2\) is mean \(\pm\) standard deviation, \(\bar{\mu}\) is the mean passivity margin, and \(t\) is the mean identification time.}
        \label{tab:fir_n21}
        \footnotesize
        \setlength{\tabcolsep}{5pt}
        \begin{tabular*}{\linewidth}{@{\extracolsep{\fill}}lccc@{}}
        \hline\hline\addlinespace[1pt]
        Method & \(E_2\) & \(\bar{\mu}\) & \(t\) [s]\\
        \hline\addlinespace[2pt]
        FSID & \(0.2473\pm0.0537\) & --- & \(0.01\)\\
        FSID-PR & \(0.3064\pm0.0930\) & \(1.198\times10^{-5}\) & \(0.22\)\\
        GN & \(0.0675\pm0.0291\) & --- & \(12.6\)\\
        GN-PR & \(0.1161\pm0.0551\) & \(6.839\times10^{-6}\) & \(12.8\)\\
        pH-Flex & \(0.1005\pm0.0272\) & \(2.496\times10^{-2}\) & \(32.0\)\\
        PR-SCP & \(0.0630\pm0.0089\) & \(6.982\times10^{-4}\) & \(65.9\)\\
        PR-SCP-H & \(0.0640\pm0.0102\) & \(1.086\times10^{-3}\) & \(13.1\)\\
        \hline\hline
        \end{tabular*}
    \end{table}

\end{document}